\documentclass[letter]{aa}
\pdfoutput=1
\usepackage{epsfig}
\usepackage{epstopdf}
\usepackage{multirow}
\usepackage[varg]{txfonts}

\def \aap  {A\&A}

\def \apjs  {ApJS}
\def \apjl  {ApJL}

\def \chisq  {\ifmmode  \chi^2   \else  $\chi^2$  \fi}  
\def \spose#1{\hbox  to 0pt{#1\hss}}  
\def \lta{\mathrel{\spose{\lower 3pt\hbox{$\sim$}}\raise  2.0pt\hbox{$<$}}}
\def \gta{\mathrel{\spose{\lower  3pt\hbox{$\sim$}}\raise 2.0pt\hbox{$>$}}}

\def \kms {\ifmmode  \,\rm km\,s^{-1} \else $\,\rm km\,s^{-1}  $ \fi }
\def \kpc {\ifmmode  {\rm~kpc}  \else ${\rm~kpc}$\fi}  
\def \pc {\ifmmode  {\rm~pc}  \else ${\rm~pc}$ \fi  }  
\def \Gyr {\ifmmode  {\rm~Gyr}  \else ${\rm~Gyr}$\fi}
\def \Msun {\ifmmode M_{\odot} \else $M_{\odot}$ \fi}
\def \Zsun {\ifmmode Z_{\odot} \else $Z_{\odot}$ \fi} 
\def \Lsun {\ifmmode L_{\odot} \else $L_{\odot}$ \fi} 
\def \Rsun {\ifmmode R_{\odot} \else $R_{\odot}$ \fi} 
\def \Msunpyr {\ifmmode M_{\odot}{\rm~yr}^{-1} \else $M_{\odot}{\rm~yr}^{-1}$ \fi} 
\def \hMsun {\ifmmode h^{-1}\,\rm M_{\odot} \else $h^{-1}\,\rm M_{\odot}$ \fi}

\def \LCDM {\ifmmode \Lambda{\rm CDM} \else $\Lambda{\rm CDM}$ \fi}
\def \sig8 {\ifmmode \sigma_8 \else $\sigma_8$ \fi} 
\def \OmegaM {\ifmmode \Omega_{\rm M} \else $\Omega_{\rm M}$ \fi} 
\def \OmegaL {\ifmmode \Omega_{\rm \Lambda} \else $\Omega_{\rm \Lambda}$\fi} 
\def \Deltavir {\ifmmode \Delta_{\rm vir} \else $\Delta_{\rm vir}$ \fi}
\def \rhocrit {\ifmmode \rho_{\rm crit} \else $\rho_{\rm crit}$ \fi}
\def \rhou {\ifmmode \rho_{\rm u} \else $\rho_{\rm u}$ \fi}
\def \zc {\ifmmode z_{\rm c} \else $z_{\rm c}$ \fi}

\def \rhos {\ifmmode \rho_{\rm s} \else $\rho_{\rm s}$ \fi} 
\def \rs {\ifmmode r_{\rm s} \else $r_{\rm s}$ \fi} 
\def \cvir {\ifmmode c_{\rm vir} \else $c_{\rm vir}$ \fi} 
\def \Rvir {\ifmmode r_{\rm vir} \else $R_{\rm vir}$ \fi}
\def \Vvir {\ifmmode V_{\rm  vir} \else  $V_{\rm vir}$  \fi} 
\def \Mvir {\ifmmode M_{\rm  vir} \else $M_{\rm  vir}$ \fi}  
\def \Nvir {\ifmmode N_{\rm  vir} \else $N_{\rm  vir}$ \fi}  
\def \Jvir {\ifmmode J_{\rm vir} \else $J_{\rm vir}$ \fi} 
\def \Evir {\ifmmode E_{\rm vir} \else $E_{\rm vir}$ \fi} 
\def \vvir {\ifmmode v_{\rm vir} \else $v_{\rm vir}$ \fi} 
\def \lam {\ifmmode \lambda  \else $\lambda$ \fi} 
\def \lamp {\ifmmode \lambda^{\prime} \else $\lambda^{\prime}$  \fi} 
\def \Vmax {\ifmmode V_{\rm  max} \else  $V_{\rm max}$  \fi} 
\def \Mdm {\ifmmode M_{\rm  dm} \else $M_{\rm  dm}$ \fi}

\def \Mgas {\ifmmode M_{\rm gas} \else $M_{\rm gas}$ \fi} 
\def \Mcg {\ifmmode M_{\rm cg} \else $M_{\rm cg}$\fi} 
\def \Mhg {\ifmmode M_{\rm hg} \else $M_{\rm hg}$ \fi} 
\def \Mdisc {\ifmmode M_{\rm disc} \else $M_{\rm disc}$ \fi} 
\def \Md {\ifmmode M_{\rm d} \else $M_{\rm d}$ \fi} 
\def \Mda {\ifmmode M_{\rm d,0\%} \else $M_{\rm d,0\%}$ \fi} 
\def \Mdb {\ifmmode M_{\rm d,20\%} \else $M_{\rm d,20\%}$ \fi} 
\def \Mdc {\ifmmode M_{\rm d,40\%} \else $M_{\rm d,40\%}$ \fi} 
\def \md {\ifmmode m_{\rm d} \else $m_{\rm d}$ \fi} 
\def \Mb {\ifmmode M_{\rm b} \else $M_{\rm b}$ \fi} 
\def \Mbh {\ifmmode M_{\rm b,pri} \else $M_{\rm b,pri}$ \fi} 
\def \Mbs {\ifmmode M_{\rm b,sat} \else $M_{\rm b,sat}$ \fi} 
\def \zo {\ifmmode z_{0} \else $z_{0}$ \fi} 
\def \rd {\ifmmode r_{\rm d} \else $r_{\rm d}$ \fi}
\def \rg {\ifmmode r_{\rm g} \else $r_{\rm g}$ \fi}
\def \rb {\ifmmode r_{\rm b} \else $r_{\rm b}$\fi}
\def \rs {\ifmmode r_{\rm s} \else $r_{\rm s}$\fi}
\def \rc {\ifmmode r_{\rm c} \else $r_{\rm c}$\fi}
\def \rvir {\ifmmode r_{\rm vir} \else $r_{\rm vir}$\fi}
\def \rbh {\ifmmode r_{\rm b,pri} \else $r_{\rm b,pri}$ \fi} 
\def \rbs {\ifmmode r_{\rm b,sat} \else $r_{\rm b,sat}$ \fi} 
\def \zp {\ifmmode z_{\rm phot} \else $z_{\rm phot}$ \fi}
\def \zs {\ifmmode z_{\rm spec} \else $z_{\rm spec}$ \fi}

\def \oii {[\ion{O}{II}]}
\def \oiil {[\ion{O}{II}] {\ifmmode \lambda\lambda\ 3726,3729\ \AA \else $\lambda\lambda\ 3726,3729$ \AA \fi}}
\def \mgii {\ion{Mg}{II}}

\begin{document}

\title{A highly-ionized region surrounding SN Refsdal revealed by MUSE} 

\author{W.~Karman\inst{\ref{inst1},\ref{mail}} \and C.~Grillo\inst{\ref{inst2}} \and I.~Balestra\inst{\ref{inst4},\ref{inst5}} \and P.~Rosati\inst{\ref{inst3}} \and K. I.~Caputi\inst{\ref{inst1}} \and E.~Di~Teodoro\inst{\ref{inst6}} \and F.~Fraternali\inst{\ref{inst1},\ref{inst6}} \and R.~Gavazzi\inst{\ref{inst7}} \and A.~Mercurio\inst{\ref{inst8}} \and J.X.~Prochaska\inst{\ref{inst9}} \and S.~Rodney\inst{\ref{inst10},\ref{inst11}} \and T.~Treu\inst{\ref{inst12}}}
\institute{ Kapteyn Astronomical Institute, University of Groningen, Postbus 800, 9700 AV Groningen, the Netherlands\label{inst1} 
\and email: karman@astro.rug.nl \label{mail}
\and Dark Cosmology Centre, Niels Bohr Institute,
  University of Copenhagen, Juliane Maries Vej 30, DK-2100 Copenhagen,
  Denmark\label{inst2}
\and INAF - Osservatorio Astronomico di Trieste, via
  G. B. Tiepolo 11, I-34143, Trieste, Italy \label{inst4}
\and University Observatory Munich, Scheinerstrasse 1, D-81679 Munich, Germany \label{inst5}
\and Dipartimento di Fisica e Scienze della Terra,
  Universit\`a degli Studi di Ferrara, Via Saragat 1, I-44122 Ferrara,
  Italy \label{inst3}
\and Department of Physics and Astronomy, University of Bologna, 6/2, Viale Berti Pichat, 40127, Bologna, Italy \label{inst6}
\and Institut d'Astrophysique de Paris, UMR7095 CNRS-Université Pierre et Marie Curie, 98bis bd Arago, F-75014 Paris, France \label{inst7}
\and INAF - Osservatorio Astronomico di Capodimonte, Via
  Moiariello 16, I-80131 Napoli, Italy \label{inst8}
\and Department of Astronomy and Astrophysics, UCO/Lick Observatory; University of California, Santa Cruz, CA 95064, USA \label{inst9}
\and Department of Physics and Astronomy, University of South
Carolina, 712 Main St., Columbia, SC 29208, USA \label{inst10}
\and Department of Physics and Astronomy, The Johns Hopkins University, Baltimore, MD 21218, USA \label{inst11} 
\and Department of Physics and Astronomy, University of California, Los Angeles, CA 90095, USA \label{inst12}
}

\abstract{
Supernova (SN) Refsdal is the first multiply-imaged, highly-magnified, and spatially-resolved SN ever observed. The SN exploded in a highly-magnified spiral galaxy at {\em z}=1.49 behind the Frontier Fields Cluster MACS1149, and provides a unique opportunity to study the environment of SNe at high {\em z}.
We exploit the time delay between multiple images to determine the properties of the SN and its environment, before, during, and after the SN exploded. 
We use the integral-field spectrograph MUSE on the VLT to simultaneously target all observed and model-predicted positions of SN Refsdal. 
We find \mgii\ emission at all positions of SN Refsdal, accompanied by weak \ion{Fe}{II*} emission at two positions. The measured ratios of \oii\ to \mgii\ emission of 10-20 indicate a high degree of ionization with low metallicity. Because the same high degree of ionization is found in all images, and our spatial resolution is too coarse to resolve the region of influence of SN Refsdal, we conclude that this high degree of ionization has been produced by previous SNe or a young and hot stellar population.  We find no variability of the \oii\ line over a period of 57 days. This suggests that there is no variation in the \oii\ luminosity of the SN over this period, or that the SN has a small contribution to the integrated \oii\ emission over the scale resolved by our observations.
}
              
\date{Received ... /
Accepted ...}


\titlerunning{Resolved MgII emission}
\authorrunning{W. Karman et al.}

\setcounter{footnote}{1}

\maketitle

\section{Introduction}
\label{sec:intro}

Supernova (SN) Refsdal \citep{Kelly2015} is the 
first observed lensed SN resolved into multiple images. Being hosted by a spiral galaxy (Sp1149) that is multiply
imaged itself, the discovery presents the possibility of studying the environment of SN Refsdal
 at three different epochs, with lensing affording spatial scales of $\sim$0.7 kpc. Preliminary strong-lensing models \citep[e.g.][]{Oguri2015,Sharon2015,Diego2015} 
predicted that we are now witnessing the second appearance of the SN. Therefore,
we are able to study the environment of the SN before, during, and after its explosion. 

Although SN Refsdal is a very interesting object, determining the properties of 
its environment could be even more important from the point of view of 
galaxy evolution. 
Galaxy growth is regulated by feedback and outflows, which are
ascribed to energetic processes such as SNe and active galactic nuclei. The 
necessity of feedback is widely acknowledged, but the detailed processes that
regulate it are still under debate. Feedback of SNe becomes more efficient when the
surrounding medium has already been heated and diluted by young and hot stars
\citep[e.g][]{Stinson2013,Hopkins2014,Artale2015,Kim2015,Martizzi2015, Vasiliev2015}, and observations of the
properties of the interstellar medium (ISM) surrounding SNe are crucial to confirm
this heating, especially at high redshift where feedback is strongest. The resolution
and luminosity required to study the environment of SNe at high redshift can currently only be achieved
through gravitational lensing.

Given its recent discovery, the environment of SN Refsdal has not yet been 
studied in detail, but its host galaxy was the focus of several works.
Sp1149 \citep{Smith2009}, located behind the Frontier Fields 
cluster MACS1149 ({\em z}=0.542), is a spiral galaxy at {\em z}=1.49 
with many star-forming clumps in its arms, and a large magnification \citep{Zitrin2009}. \citet{Yuan2011,Yuan2015} found a 
steep metallicity gradient in Sp1149, measuring a low metallicity 
({\em Z} $\lesssim$0.3$Z_\odot$) at the position
of SN Refsdal. Active, widespread star formation (SF) was shown by rest-frame UV 
observations \citep{Smith2009}, and through \ion{H}{$\alpha$} detection 
\citep{Livermore2012,Livermore2015} at a rate of a few \mbox{\Msun yr$^{-1}$}. The 
size and age of the clumps increase inwards, with a young population of stars
at the position of SN Refsdal \citep{Adamo2013}.

\begin{figure*}
\begin{center}
  \includegraphics[width=.95\textwidth]{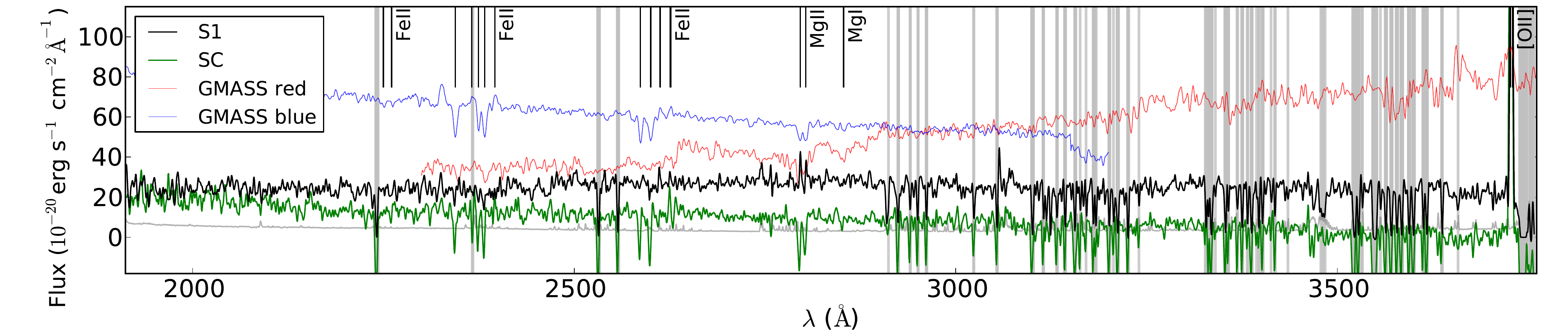}
\caption{ Rest-frame spectrum of Sp1149 at the position S1 and its centre (SC),
which have been smoothed over 9 pixels for illustrative purposes.
For comparison, the red and blue composite GMASS spectra \citep{Kurk2013} are also 
shown, which are normalized to the Sp1149 spectrum at 3000 \AA. To avoid confusion, we added an offset of 
(-25,  25) $\times 10^{-20}$ erg s$^{-1}$ cm$^{-2}$ \AA$^{-1}$ to the central spectrum and the
GMASS spectra, respectively. The error estimates are shown by
the grey line, while the vertical grey bands are located at wavelengths 
with significant skyline contamination. The short black vertical lines
show the wavelength of important \ion{Fe}{} and \ion{Mg}{} lines.
\label{fig:spectrum}} 
\end{center}
\end{figure*}

The environment of SN Refsdal can be studied through the presence of certain
emission lines. In the optical regime, \ion{Mg}{II} is frequently used, as it
is detectable from the ground over a large redshift range.
\citet{Weiner2009} were the first
to report \mgii\ emission lines in galaxies at $z>1$, and these lines
have been found in several studies since 
\citep[][]{Pettini2010,Rubin2011,Martin2012,Erb2012,Kornei2013, Rigby2014,Tang2014,Zhu2015}. 
The origin of \mgii\ emission can either be \ion{H}{II} regions 
\citep[see][]{Kinney1993}, or resonant scattering 
in outflowing gas \citep{Prochaska2011}. \citet{Rubin2011} found \mgii\ 
emission in a galaxy at $z=0.69$ that was clearly extended further out than the continuum emission,
in support of the latter scenario. However, \citet{Erb2012} found that the
\mgii\ emission in their sample was more similar to \ion{H}{II} regions, 
supporting the former.

In this paper, we study the properties of the region surrounding SN Refsdal in the highly-magnified, multiply-imaged, spiral galaxy Sp1149 at {\em z=1.49}
through high-quality integral field spectroscopy, at all of its observed and model-predicted multiple image positions. 
In Sect.~\ref{sec:data} we present
the observations, in Sect.~\ref{sec:results} we show the results of our
observations, we discuss the implications in Sect.~\ref{sec:discussion}, 
and we show our conclusions in Sect.~\ref{sec:conclusion}.


\section{Observations}
\label{sec:data}

We observed MACS1149 with the  Multi Unit Spectroscopic Explorer 
\citep[MUSE,][]{Bacon2012} instrument mounted on the Very Large Telescope at 
Paranal observatory. The total observation time was 6 hours of the Director's 
Discretionary Time (DDT), and were executed in service mode (PI:~C.~Grillo; ID~294.A-5032).
The telescope was centered at $\alpha$=11:49:35.75, $\delta$=22:23:52.4, such that
three images of Sp1149 are within the field of view (FOV).
The observations were carried out in the nights of 
February 14th 2015 (1 hr), March 21st 2015 (4 hrs), and April 12th 2015 (1 hr).
 Each hour consisted of two exposures of 1440 seconds each, such 
that the total exposure time adds up to 17280 seconds or 4.8 hours. The exposures
were taken under clear and photometric conditions, and the seeing was 
$<$1.1$\arcsec$ in 10 out of the 12 exposures. The last hour of the  
observations in March was executed with a significantly worse seeing of 
$\sim$2$\arcsec$.

The data reduction was performed using the MUSE Data Reduction Software version
1.0.
The final datacube consists of pixels with a spatial
scale of 0.2$\arcsec\times$0.2$\arcsec$ and a spectral scale of 1.25~\AA,
 although we also constructed a datacube with a higher spectral sampling
of 0.82~\AA. 
We measured the spatial full width half maximum (FWHM) of the final datacube
from a bright star in the FOV to be $0.9\arcsec$, while the spectral resolution
is $\sim$2.3~\AA. This seeing corresponds to a spatial scale in the source plane of
$\sim$500 pc for a maximum magnification of $mu=16$ or 4 kpc for the minimum 
magnification $\mu=2$ in Sp1149. For details on the data reduction, see \citet{Karman2015}.

We extracted individual spectra in $0.6\arcsec$-radius apertures, based on the 
SN coordinates from {\em Hubble Space Telescope}~({\em HST}) observations and strong-lensing-modeling predictions 
\citep[S1-4, SX, SY, following][]{Oguri2015}. We updated
the model-predicted positions of SX and SY according to our refined strong-lensing model 
(Grillo et al. in prep), and show all positions and magnification factors
$\mu$ in Table~\ref{tab:locs}. We find no change in our conclusions when we vary the
positions within the uncertainties of our model, or by adopting positions predicted by other authors.
The spectroscopic
observations span approximately the rest-frame wavelength range of $1910$ \AA\ through 3740 \AA,
which includes the spectral lines \ion{Fe}{II}, \ion{Mg}{II}, and [\ion{O}{II}]. 

\section{Spectral analysis}
\label{sec:results}

\begin{table}
 \begin{center}
 \begin{tabular}{cccccc}
\hline \hline
  Cl. & R.A. & Decl. & $\mu^{(1)}$ & F$_{\ion{O}{II}}^{(2)}$ & F$_{\ion{Mg}{II}}^{(3)}$ \\
        & (J2000) & (J2000) & & \multicolumn{2}{c}{(10$^{-18}$ erg s$^{-1}$cm$^{-2}$)} \\
\hline
 S1 & 177.39821 & 22.39563 & 13.5$^{+3.9}_{-2.2}$ & 39.1$\pm1.7$ & 2.6$\pm0.6$\\
 S2 & 177.39771 & 22.39579 & 12.4$^{+6.4}_{-3.5}$ & 29.2$\pm1.0$ & 2.3$\pm1.0$ \\
 S3 & 177.39737 & 22.39554 & 13.4$^{+5.8}_{-3.1}$ & 20.4$\pm0.9$ & 1.6$\pm1.1$\\
 S4 & 177.39780 & 22.39517 & 5.7$^{+2.4}_{-2.0}$ & 16.3$\pm1.2$ & 1.8$\pm1.3$ \\
 SX & 177.40012 & 22.39670 & 4.8$^{+0.5}_{-0.3}$ & 15.0$\pm1.9$ & 1.6$\pm0.8$\\
 SY & 177.40382 & 22.40205 & 4.0$^{+0.2}_{-0.2}$ & 15.2$\pm1.7$ & 1.5$\pm1.2$\\
 SC & 177.39702 & 22.39599 & 7.0$^{+0.6}_{-0.5}$ & 56.8$\pm1.1$ & --- \\
 F1 & 177.39744 & 22.39641 & 11.7$^{+1.3}_{-0.8}$ & 25.8$\pm1.5$ & --- \\
 F2 & 177.39883 & 22.39769 & 8.5$^{+0.8}_{-0.6}$ & 26.7$\pm1.9$ & 2.9$\pm1.4$ \\
 F3 & 177.39975 & 22.39656 & 9.4$^{+0.7}_{-0.8}$ & 19.6$\pm1.6$ & 2.1$\pm1.0$ \\
 F4 & 177.40313 & 22.40257 & 3.9$^{+0.1}_{-0.2}$ & 30.5$\pm1.8$ & 0.2$\pm0.5$ \\
\hline
 \end{tabular}
\end{center}
\caption{Observed (S1-4) and model-predicted (future: SX, past: SY) positions of SN Refsdal, the 
centre of Sp1149 (SC), and 4 star-forming clumps for comparison (F1-4). $^{(1)}$ median 
values of the model-predicted magnification factor (Grillo et al. in prep), 
$^{(2)}$ emission line flux of \oii, $^{(3)}$ emission line flux of
\mgii~$\lambda$~2796~\AA\ corrected for stellar absorption \citep[see][]{Guseva2013}. 
The emission line fluxes are deteremined in 
an 0.6\arcsec-radius aperture, and 
are not corrected for magnification.
\label{tab:locs}}
\end{table}

\begin{figure*}
\begin{center}
 \includegraphics[width=.46\textwidth]{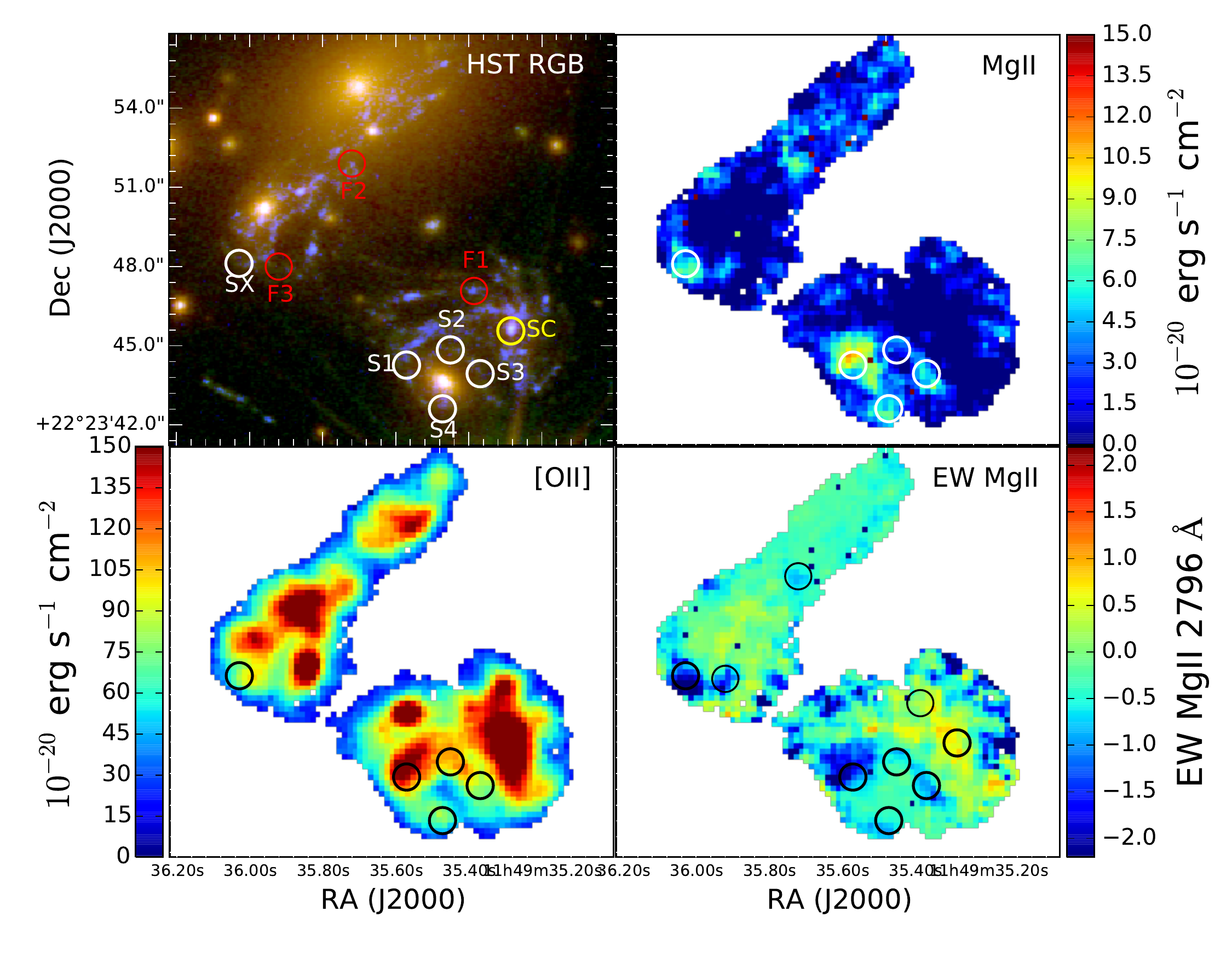}
 \includegraphics[width=.46\textwidth]{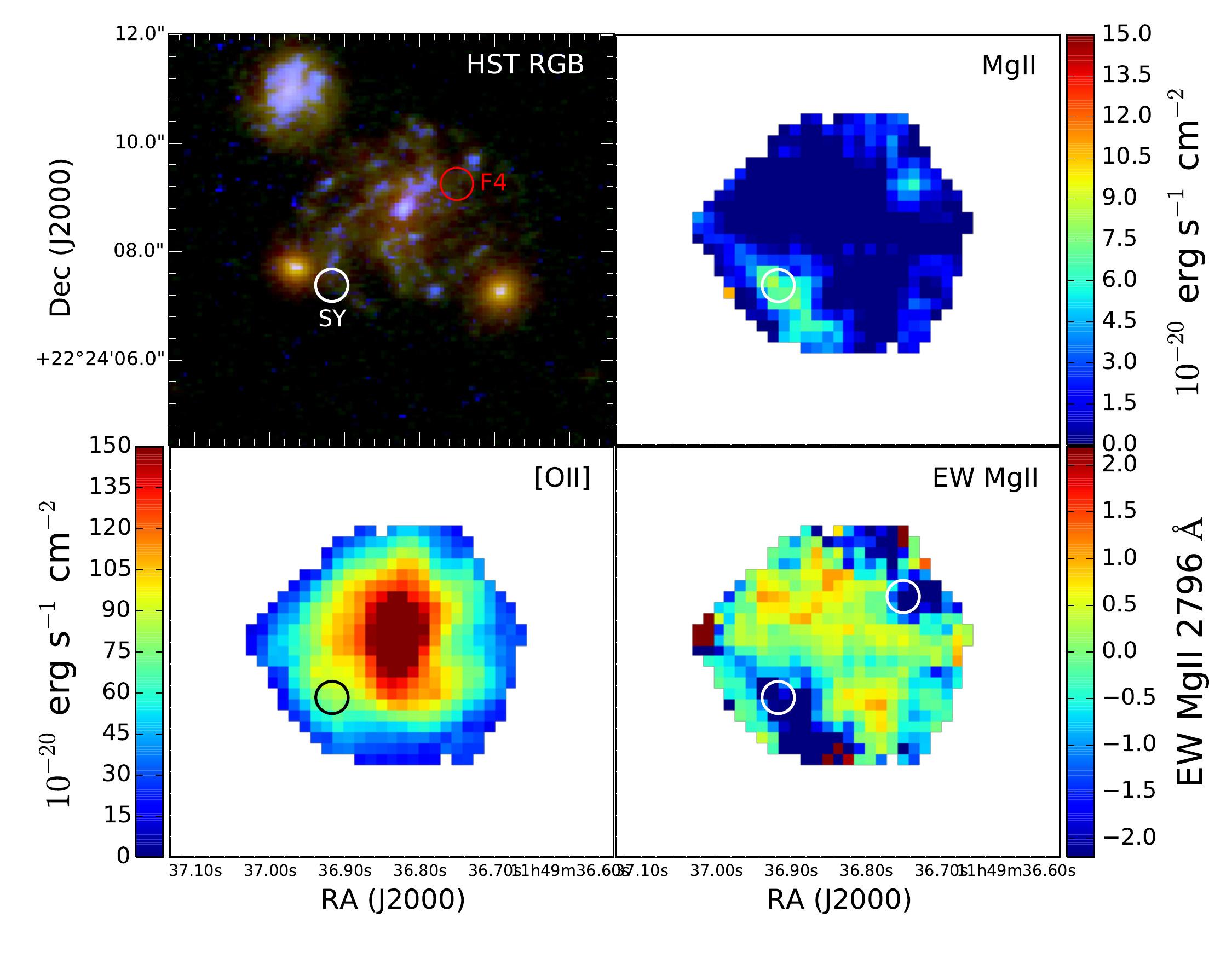}
\caption{ The two western images (left), and the eastern image (right) of 
Sp1149. The two figures are divided into four panels; 
{\em top left:} {\em HST} RGB image with 
white, yellow, and red circles at the positions of SN Refsdal, the centre of the galaxy, and four star-forming clumps of Sp1149, respectively. 
{\em top right:} 
\mgii\ emission line flux, {\em bottom left:} \oii\ emisson-line flux {\em bottom right:} equivalent width (EW) of 
\mgii\ $\lambda 2796$ \AA, where a negative EW represents emission. 
The circles are at the same positions as the circles in the top left panel.
The EW have been corrected for stellar absorption \citep[0.5 \AA, see][]{Guseva2013},
and, for displaying purposes, the MUSE maps have been smoothed over a box of 
3$\times$3 pixels. 
\label{fig:mg2map}} 
\end{center}
\end{figure*}

We show the rest-frame spectrum at S1 in Fig.~\ref{fig:spectrum}, where
 we compare the spectrum to that in the centre of the
galaxy (SC) and a blue-galaxy composite spectrum from the Galaxy 
Mass Assembly ultra-deep Spectroscopic Survey \citep[GMASS][]{Kurk2013}.
The slopes of the spectra at S1 and SC are very similar, and comparable
to the composite of GMASS. This is not surprising, as different
studies have shown spread-out SF in Sp1149 \citep[e.g.][]{Livermore2012}.
The slight difference in slope between S1 and SC is probably caused by
contamination of S1 from the foreground red cluster galaxy which is 
responsible for the Einstein cross. 
Looking at the individual 
lines, the strong \oii\ emission doublet is clear, and some other
lines show important differences between S1 and the other spectra. While the spectrum at SC and the blue 
composite GMASS spectrum show prominent absorption lines, these lines are
significantly weaker or absent at S1. This is clearest at the 
\ion{Fe}{II} $\lambda\lambda$ 2587, 2600~\AA, the 
\mgii~$\lambda\lambda$~2796,~2803~\AA, and the \ion{Mg}{I}$\lambda$~2853~\AA\ 
absorption lines. 

\begin{figure}
\begin{center}
 \includegraphics[width=.75\columnwidth]{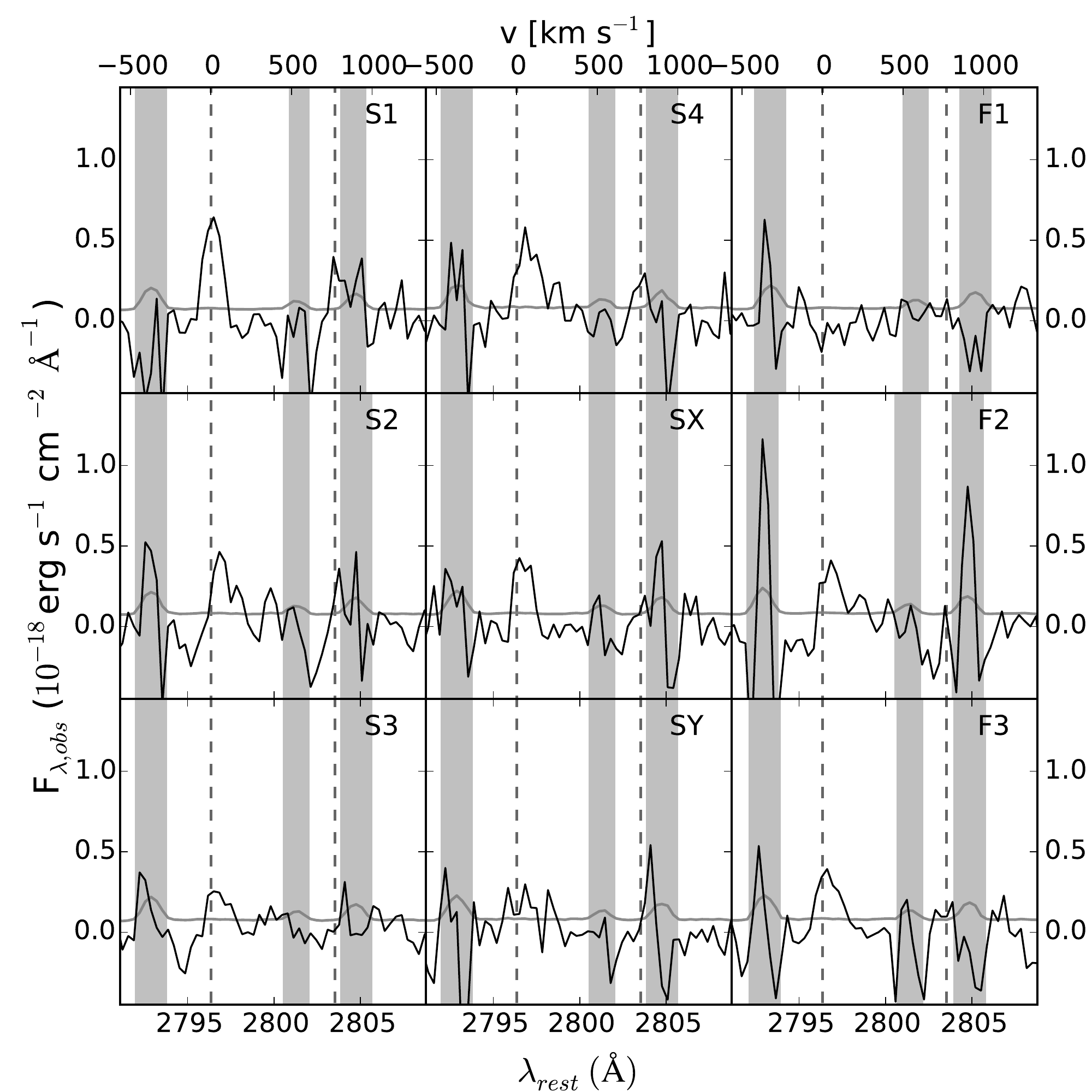}
\caption{ Spectra around the \mgii\ line at all observed and model-predicted SN
positions (S1-4,SX,SY), and at three star forming clumps (F1-3). The black
solid lines show the continuum-subtracted spectra which have been shifted to their rest frame using
the \oii\ doublet to determine the redshift. 
The vertical dashed lines are at the wavelengths
of the \mgii\ doublet, the grey solid lines represent the errors, and grey vertical bands mark
wavelengths with significant sky emission. 
The axis on the top shows the relative velocity of \mgii~$\lambda$~2796\AA\ with respect to \oii.
\label{fig:mgiispectra}} 
\end{center}
\end{figure}

While \mgii\ absorption is not seen at any of the SN positions, 
\mgii\ emission is visible at all of them, see Figs. \ref{fig:mg2map}, \ref{fig:mgiispectra} and Table~\ref{tab:locs}.
Although the majority of previously reported \mgii\ emitters also show clear \mgii\ 
absorption, some studies found weak or no accompanying absorption 
\citep[][]{Weiner2009,Erb2012,Guseva2013,Rigby2014}.
In our case, the absent \mgii\ absorption, the similarly small rest-frame FWHM of $\sim$120 and $\sim$140 km~s$^{-1}$ for
\mgii\ and \oii\ respectively,
and the negligible velocity offset of \mgii\ compared to \oii, indicate that the \mgii\ emission originates
from \ion{H}{II} regions, rather than from resonant scattering.

In Fig.~\ref{fig:mg2map}, we show maps of \oii\ and \mgii\ emission throughout 
Sp1149 in all three images. While the \oii\ emission is widespread, the \mgii\ emission is local.
 The clearest \mgii\ emission we see is at position S1,
while the center of the galaxy shows strong absorption of \mgii. The measured lower fluxes
 at positions S2-4 are probably due to a lower $\mu$ when
integrated around S2-4, which differs from the values of point-like images in 
Table~\ref{tab:locs}. Since we are interested in the ISM, we correct the equivalent width and
flux of \mgii\ for stellar absorption \citep[0.5 \AA\ for young stellar populations, for details, see][]{Guseva2013}.

In Fig.~\ref{fig:mg2map} we see three additional regions, F2, F3, and F4, with significant flux from
the \mgii\ line, which are located too far from the SN to be influenced by it.
After extracting spectra of these regions, we find that the
spectral properties at F2 and F3 are very similar to those of S1, see Fig.~
\ref{fig:mgiispectra} for a comparison of the \mgii\ lines. At F4, we find no
significant detection of \mgii, in an aperture of $0.6\arcsec$, see Table~\ref{tab:locs}.
However, we find a weak emission line accompanied by more significant absorption 
when the size of the aperture is reduced. From the {\em HST}
images we can identify F2 and F3 with two other star forming clumps
in Sp1149. Extracting spectra of other SF clumps does not reveal more
\mgii\ emitting regions, which shows that the ionization conditions vary
significantly between clumps in Sp1149. We note that although previous work
showed that the properties of clumps, such as metallicity and stellar age, vary
significantly, these variations cannot explain the observed large differences of
the line ratios.

It is worth noticing that we detect \mgii\  emission also at position SX, which is where the SN has
 not exploded yet due to a longer time delay. This indicates that the high 
degree of ionization is not caused by the SN
, but that it
was already present before the SN went off. This is strengthened by the
detection of additional SF clumps with \mgii\ emission. 
We use the ratio of \oii\ and
\mgii\ flux to estimate the ionization parameter {\em U}, following \citet{Erb2012}.
The emission line flux is determined by fitting a Gaussian to the spectrum,
while the uncertainties are determined by repeating the calculation on 
1000 mock spectra. The mock spectra are created by applying a scatter 
randomly picked from a normal distribution with a standard deviation equal to 
the error at every spectral element.  
We measure a ratio of \mbox{10-20} for S1-4, SX, and SY, where the main differences
are due to the relatively low S/N and nearby skylines. The models of
\citet{Erb2012} predict an ionization parameter $-2<\log(U)<-1$ for such ratios,
which is higher than that found for \ion{H}{II} regions \citep[$-3<\log(U)<-2$,][]{Kinney1993},
and a low metallicity in agreement with the results of \citet{Yuan2015}. 
A metallicity even lower than used by \citet{Erb2012}, i.e. $Z<0.2\Zsun$, could
possibly reconcile the measured ratio with the ionization parameter found in 
\ion{H}{II} regions, although we note that our measured ratio is already in agreement
with the ratios found by \citet[][]{Guseva2013}, i.e. \oii/\mgii=5-20.

We find weaker and narrower \ion{Fe}{II*} $\lambda\lambda$ 2612,2626 \AA, at S1
and S2. This non-resonant emission is accompanied by \ion{Fe}{II} absorption 
lines that are weaker than observed in the rest of the galaxy. Due to the 
proximity of sky lines and lower magnification factors, we are unable to 
confirm or reject the  presence of these lines at S3, S4, SX, and SY.

\begin{table}
 \begin{center}
 \begin{tabular}{cccc}
\hline \hline
  Pos. & Feb. 14th & Mar. 21st & Apr. 12th \\
        & \multicolumn{3}{c}{(10$^{-18}$ erg s$^{-1}$cm$^{-2}$)} \\
\hline
 S1 & 38.2$\pm3.9$ & 41.7$\pm2.3$ & 40.5$\pm5.6$ \\
 S2 & 28.0$\pm4.7$ & 28.6$\pm2.2$ & 32.6$\pm6.5$ \\
S3 & 22.4$\pm4.1$ & 22.4$\pm2.6$ & 29.2$\pm6.3$ \\
S4 & 16.3$\pm6.6$ & 16.3$\pm2.1$ & 15.6$\pm5.5$ \\
\hline
 \end{tabular}
\end{center}
\caption{ Variability of the \oii\ emission line at the four observed positions
of SN Refsdal (S1 to S4). 
The emission line fluxes have been measured by fitting a 
Gaussian to the emission. \label{tab:oiivar}}
\end{table}

The \oii\ line offers an opportunity to study a possible increase in brightness of the SN. Since our data were taken in three separate months, we can compare the \oii\ fluxes at intervals of 35 and 57 days. We estimate the spatial resolution of our observations to be $\sim500-700$ pc at the positions of the SN, using the determined FWHM and the magnification factors presented in Table \ref{tab:locs}. This spatial resolution is less than the resolution of {\em HST}, and much larger than the region of influence of the SN. However, the contrast between the SN region and the lensing cluster galaxy is much better at the wavelengths of the  \oii\ line than in the continuum, and if the SN were the dominant source of \oii\ emission in this 500 pc region, we should detect the variability. We study the \oii\ line at the three different epochs, and find no significant change in the \oii\ luminosity or peak position at the individual positions, see Table~\ref{tab:oiivar}. The lack of variability means 
either no variation in the SN \oii\ luminosity over this period, or a relative contribution of the SN to the \oii\ line that is too small to distinguish with our spectrophotometric precision. The small variation is not surprising, as only SN with ejecta interacting with circumstellar matter are observed to have clear \oii\ emission in their spectra \citep[e.g.][]{Filippenko1997,Smith2009SN,Pastorello2015}.

%
\section{Discussion}
\label{sec:discussion}

Most other galaxies where \mgii\ emission is found in absence of absorption 
are also lensed by foreground clusters and galaxies 
\citep[e.g.][]{Pettini2010,Rigby2014}. Similar to these, we find
that only some regions of Sp1149 show \mgii\ emission, while other regions
show clear absorption. This means that if the complete galaxy is observed with a single
slit, the absorption line would dominate the spectrum. This highlights
the importance of spatially resolving galaxies, which is currently only possible at $z>1$ 
for gravitationally-lensed and magnified sources.

The \mgii\ emission accompanied by a weak \ion{Fe}{II}$\lambda$~2383\AA\ 
absorption line, is predicted by the models of \citet{Prochaska2011}
and observed by \citet{Erb2012}. While most \ion{Fe}{} absorption lines are coupled to 
fine structure transitions, \ion{Fe}{II}$\lambda$~2383\AA\ is only able to 
fall back to the ground state. This leads to a larger degree of emission 
line filling, similar to \mgii. However, this also suggests that the other
\ion{Fe}{} transitions should be clearly visible in the spectra, and it remains unclear
why recombination in the \ion{H}{II} regions would affect the resonant \ion{Fe}{II} lines differently than
the resonant \mgii\ lines. 

The absence of the \ion{Fe}{II}
lines at all positions of SN Refsdal indicates that the medium is optically 
thin, at least for these transitions, and that this is not caused by the 
exploding SN. 
The low opacity at the \ion{Fe}{} lines agrees with the
occurence of \mgii\ emission from \ion{H}{II} regions, as that would also require an optically thin
gas in front of the emitting regions, possibly leaking continuum-ionizing-photons.
 Assuming $\tau_{\ion{Mg}{II}}<1$ and 
$Z=0.2\Zsun$ limits the column density to $N_{\rm H}<3 \times 10^{16}$ 
cm$^{-3}$. This provides another indication that the feedback of hot young 
stars, possibly in combination with previous SNe, has removed most of the cold 
gas from the environment before SN Refsdal exploded, without forming the shell
required for resonant scattering. The early removal of cold gas is in line with
predictions of recent feedback models, where young and hot stars dilute the ISM before 
SN explosions, and therefore provides a first step towards the 
confirmation of early-stellar feedback.

\section{Summary and conclusions}
\label{sec:conclusion}

We use VLT/MUSE to study the properties of the ISM around the multiply-imaged SN Refsdal. We find 
clear \mgii\ and weak \ion{Fe}{II*} emission, in combination with almost
absent \ion{Fe}{II} absorption. We determine that the \mgii\ emission is constrained
to the clump surrounding the SN, and also find it at the model-predicted positions
in the other two images. This shows that these conditions are not caused by the
SN, but are rather the consequence of the young stellar population as a whole. 
By using the ratio of the \oii\ and \mgii\ flux, we show that the ionization  
parameter is higher than normally seen in \ion{H}{II} regions. This can 
have important consequences for feedback models, as our results suggest to also include stellar feedback.

We use the \oii\ lines to 
detect possible variation in the luminosity of the SN, and thereby constrain 
the light curve. However, we do not find a significant variability, 
suggesting that the SN is either not variable on these timescales, or it is not a 
major contributor to the  integrated \oii\ flux at the scale resolved by our 
observations.

\begin{acknowledgements}
 
Based on observations made with the European Southern Observatory Very Large Telescope (ESO/VLT) at Cerro Paranal, under program ID 294.A-5032. The authors thank Patrick Kelly and Pece Podigachoski for useful comments and discussions. This research made use of APLpy, an open-source plotting package for Python hosted at http://aplpy.github.com.

\end{acknowledgements}

\bibliographystyle{aa}
\bibliography{muse.bib}

\end{document}